\documentclass[epj]{svjour}

\usepackage{dcolumn}
\usepackage{bm}
\usepackage{footnote}
\makesavenoteenv{table*}
\makesavenoteenv{tabular}
\usepackage{graphicx}
\usepackage[square,numbers,sort,compress]{natbib}
\usepackage{amsmath}
\usepackage{amssymb}
\usepackage{gensymb}
\usepackage{hyperref}

\begin{document}

%\linenumbers

\title{Are the Muonic Hydrogen and Electron Scattering Experiments
Measuring the Same Observable?}
\titlerunning{Are Muonic Hydrogen and Electron Scattering Measuring the Same
Observable ?}

\author{T.W.~Donnelly, D.K.~Hasell, R.G.~Milner}
\institute{Center for Theoretical Physics, Laboratory for Nuclear
  Science, and Department of Physics\\Massachusetts Institute of
  Technology, Cambridge, MA 02139}
\mail{R.G.~Milner \email{milner@mit.edu}}
\authorrunning{Donnelly, Hasell, and Milner}
\date{\today}

\abstract{Elastic scattering of relativistic electrons from the
  nucleon yields Lorentz invariant form factors that describe the
  fundamental distribution of charge and magnetism.  The spatial
  dependence of the nucleon's charge and magnetism is typically
  interpreted in the Breit reference frame which is related by a
  Lorentz boost from the laboratory frame, where the nucleon is at
  rest.  We construct a model to estimate how the Sachs electric 
  and magnetic form factors can be corrected for the effects of 
  relativistic recoil.  When the corrections are applied, the ratio 
  of the proton's Sachs form factors is approximately flat with $Q^2$, i.e. the 
  spatial distributions of the proton's intrinsic charge and magnetization are similar.
  Further, we estimate the correction due to recoil that 
  must be applied to the determination of the proton charge radius
  from elastic electron scattering before it can be compared to the value determined using the Lamb shift in hydrogen.  Application of the correction brings the two values of the proton charge radius into significantly closer agreement.
  Predicted corrections based on the model are provided for the 
  rms charge radii of the deuteron, the triton, and the helium isotopes. 
\PACS{{14.20.Dh}{} \and {13.40.Gp}{} \and {25.30.Bf}{}}
\keywords{proton radius -- proton form factors -- elastic electron
scattering -- Lamb shift -- few body nuclei}
}
\maketitle

Of great current interest is the {\it proton radius puzzle}, namely that
the charge radius of the proton as determined
from precision elastic electron-proton
scattering~\citep{Bernauer:2010wm} disagrees with a high precision
determination obtained from the Lamb shift in muonic
hydrogen~\citep{Pohl:2010zza,Antognini:1900ns}.  This 
discrepancy has sparked considerable
interest~\citep{Pohl:2013yb,Bernauer:2014cwa}.  In this Letter we
question the fundamental and widely-accepted {\it ansatz} that they are
measuring the same quantity.  
%We have previously written up these arguments
%from a more general perspective~\citep{Donnelly:2015tt}.

Quantum electrodynamics describes the energy levels of hydrogen (both electronic and muonic) with great accuracy.
In particular, the energies of $S$-states in hydrogen are given by%
\begin{equation}
E(nS) \approx -\frac{R_{\infty}}{n^2} + \frac{L_{1S}}{n^3} \ ,
\end{equation}% 
where $n$ is the principal quantum number and $L_{1S}$ denotes the Lamb shift of the 
1$S$ ground state, which depends on the proton charge radius $r_p$.   For electronic hydrogen, 
$L_{1S} \approx (8,712 + 1.56 r^2_p)$ MHz when $r_p$ is expressed in femtometers.
Thus, the
finite size effect is of order 1.2 MHz.  We emphasize that this analysis of the hydrogen energy levels
is carried out in {\bf coordinate space}, {\it i.e.} the hydrogen atom is solved exactly (fully relativistically) using the Dirac equation.  We note that recent work~\citep{Beyer2017} in hydrogen spectroscopy reports a value of the Rydberg constant in tension with the world's data.

Recent technical advances have made stopped muon beams of unprecedented intensity and quality available.   These
have allowed precision measurements of the Lamb shift in muonic hydrogen which have yielded a determination
of the proton charge radius of 0.84087(39) fm.   

Consider relativistic elastic electron scattering from the nucleon.
In single photon exchange approximation, the unpolarized elastic eN
scattering cross section in the lab. system (nucleon at rest) can be
written
\begin{equation}
\frac{d\sigma }{d\Omega }=\sigma _{M}f_{rec}^{-1}\cdot \frac{1}{(1+\tau ) \epsilon} \left[ \epsilon G_{E}^{2}+\tau G_{M}^{2} \right],  
\end{equation}
where the Mott cross section is%
\begin{equation}
\sigma _{M}=\left[ \frac{\alpha \cos \theta /2}{2 E_e \sin ^{2}\theta /2} %
\right] ^{2} \ ,  \label{sigmamott}
\end{equation}%
$G_E(Q^2)$ and $G_M(Q^2)$ are the nucleon electric and magnetic Sachs form factors, $\epsilon^{-1}=1+2(1+\tau )\tan ^{2}\theta /2$, $\tau =Q^{2} / 4m_{N}^{2} \ge 0$ and the recoil factor is $f_{rec}=E_e / E_e^{\prime }$.  Here the
incident electron has energy $E_e $, the scattered electron has energy
$E_e^{\prime }$ and the scattering is through angle $\theta $; also
$\alpha $ is the fine-structure constant. In all expressions, the
extreme relativistic limit is taken, namely the electron mass is
ignored with respect to its energy.

In analysis of elastic electric-proton scattering data, the proton charge radius (denoted $r^{scatt}_{E,p}$) is determined
through the derivative of $G_{E}^{p}(Q^{2})$ with respect to the
invariant 4-momentum transfer, $Q^{2}$, namely via%
\begin{eqnarray}
\left[ -6\frac{dG_{E}^{p}(Q^{2})}{dQ^{2}}\right]_{Q^2 = 0}\equiv \left( r^{\mathrm{scatt}}_{E,p} \right) ^{2}. \label{eqadd1}
\end{eqnarray}%
Since both $G_{E}^{p}(Q^{2}) $ and $Q^{2}$ are Lorentz invariants, the
quantity $r^{\mathrm{scatt}}_{E,p}$  is also, and we point out that
this type of proton charge radius radius is determined in {\bf momentum space}.
%It is a convenient quantity to employ when
%inter-comparing data on elastic $ep$ scattering. 
It is not the RMS charge radius of the proton, however, which would be
determined by taking the charge distribution of the proton in its rest
frame, weighting by $r^{2}$, integrating and taking the square root.
% to
%obtain the coordinate-space charge radius $r^{\mathrm{coord}}_{E,p}$. 

The central point of this paper is to question whether the proton charge radius, measured
via the Lamb shift in muonic hydrogen  ($r_p$ above) in coordinate space, is different from that extracted via electron
scattering $r^{\mathrm{scatt}}_{E,p}$ in momentum space.  While $r_p$ is determined using an essentially static proton,
$r^{\mathrm{scatt}}_{E,p}$ involves a process where the proton must recoil after absorbing the momentum transfer from the exchanged virtual photon. 

We stress that this last point was recognized in the earliest work on electron scattering. 
For example, Yennie {\it et al.} pointed out~\citep{Yennie:1957aa} in 1957 that there would be a $Q^2$ dependence of the
form factors determined in electron scattering which was independent of structure and ``which would be kinematic in origin."  Fundamentally,  this arises from the relativistic recoil which is unavoidable and means that ``intuitive concepts of 
static charge and current distributions are no longer valid"~\citep{Yennie:1957aa}.  We will show that, at the precision demanded by the proton charge radius comparison, these kinematic effects are non-negligible.

Consider the Breit reference frame, defined by zero energy transfer of the
virtual photon and reversal of the 3-momentum of the target between
the initial and final states.  Thus, one has $\omega _{B} =0$ and
$q_{B} =\sqrt{|Q^{2}|}$.  Since in that frame (with $z$-axis along the
momentum transfer vector) one has the nucleon entering with 3-momentum
$-p_{B}$ and leaving with 3-momentum $+p_{B}$, one has
$p_{B}=q_{B}/2$.  Thus, the relativistic $\gamma $-factor relative to
the lab. frame for the nucleon in that frame is%
\begin{equation}
\gamma =\sqrt{1+\tau }.  \label{eq23}
\end{equation}
We note that, no matter what reference frame is adopted, the nucleon
before and after the scattering must have different momenta.

We now formulate a non-relativistic model of eN
scattering. Assume that one puts a single nucleon into the lowest
level in a very deep harmonic oscillator (HO) potential to avoid any
recoil problem. In this model, the nucleon is bound and held essentially at rest.
Physically, this is similar to the case of a nucleon bound in a heavy nucleus and not allowed to
carry the full recoil momentum when an electron scatters from it.  Using,
the tables of \citep{Donnelly:1979ezn} or the review
article of \citep{DeForest:1966ycn}, one can compute the multipole
matrix elements of the C0 (Coulomb monopole) and M1 (magnetic dipole)
elastic scattering operators.  
%As discussed in the standard literature
%for electron scattering from nuclei, and those papers in particular,
%the differential cross section may be written as in eq.~\ref{dsigma}
%and~\ref{sigmamott} where, 
Using the non-relativistic limit for the
current operators, together with $1s_{1/2}$ harmonic oscillator wave
functions, one obtains the following:%
\begin{eqnarray}
 \left\langle 1s_{1/2}\left\Vert M_{0}^{Coul} \right\Vert
1s_{1/2}\right\rangle  &=& F_{1}\left\langle 1s_{1/2}\left\Vert
M_{0}\right\Vert 1s_{1/2}\right\rangle  \label{eq24} \\
\left\langle 1s_{1/2}\left\Vert  iT_{1}^{mag} \right\Vert
1s_{1/2}\right\rangle  &=& \frac{q}{m_{N}} [ F_{1}\left\langle
1s_{1/2}\left\Vert \Delta _{1}(q\mathbf{x})\right\Vert 1s_{1/2}\right\rangle \nonumber \\
-\frac{1}{2}\left( F_{1}+F_{2}\right) &\cdot& \left\langle 1s_{1/2}\left\Vert \Sigma
_{1}^{\prime }(q\mathbf{x})\right\Vert 1s_{1/2}\right\rangle ] \ . 
\label{eq25}
\end{eqnarray}
Here, following \citep{DeForest:1966ycn}, $F_1$ and $F_2$ are the Dirac and Pauli
single-nucleon form factors.  Either working directly
with the harmonic oscillator wave functions and the explicit forms for
the current multipole operators (see \citep{DeForest:1966ycn}) or, 
using the tables of \citep{Donnelly:1979ezn}, one finds for the three
required reduced matrix elements
\begin{eqnarray}
\sqrt{4\pi }\left\langle 1s_{1/2}\left\Vert M_{0}(q\mathbf{x})\right\Vert
1s_{1/2}\right\rangle  &=&\sqrt{2}e^{-y} \\
\sqrt{4\pi }\left\langle 1s_{1/2}\left\Vert \Delta _{1}(q\mathbf{x}%
)\right\Vert 1s_{1/2}\right\rangle  &=&0 \\
\sqrt{4\pi }\left\langle 1s_{1/2}\left\Vert \Sigma _{1}^{\prime }(q\mathbf{x}%
)\right\Vert 1s_{1/2}\right\rangle  &=&2e^{-y}.
\end{eqnarray}
Each is proportional to $\exp (-y)$ where $y=(bq/2)^{2}$ with $b$ the
HO parameter. However, one must multiply by the center-of-mass
correction which in the non-relativistic HO shell model can be
computed: it is a multiplicative factor of
$f_{cm}=\exp (+y/A)=\exp (+y)$ for $A=1$ and cancels the above factor,
leaving only the remaining factors obtained by using the above-cited
tables. 

Using this model, one can then correct the Sachs form factors for the
effect of recoil to obtain elastic electric and magnetic form factors ($G^{int}$) that
are dominated by the {\it intrinsic} charge and magnetic structure
\begin{eqnarray}
G_{E}^{int} &\equiv &\sqrt{1+\tau }G_{E}  \label{eq27} \\
G_{M}^{int} &\equiv &\frac{1}{\sqrt{1+\tau }}G_{M} \ .  \label{eq28}
\end{eqnarray}%

\begin{figure}[!ht]
\centering
\includegraphics[width=0.48\textwidth, viewport=0 188 559 569, clip]
{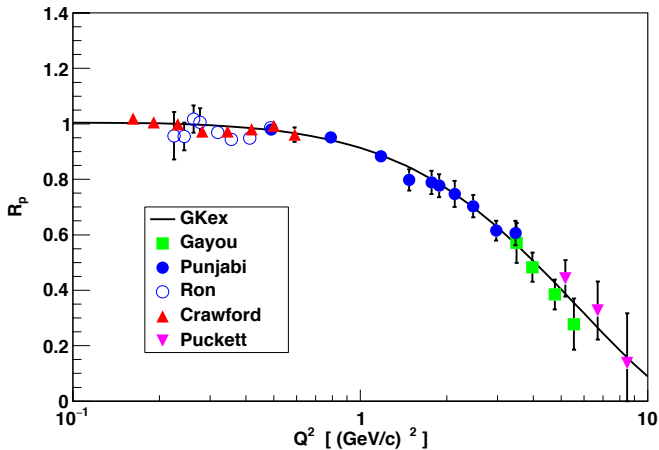}
\caption{Plot of $R_p$, defined in Eq.(\ref{eq29}), versus $Q^2$.}
\label{fig:1}
\end{figure}

\begin{figure}[!ht]
\centering
\includegraphics[width=0.48\textwidth, viewport=0 188 595 569, clip]
{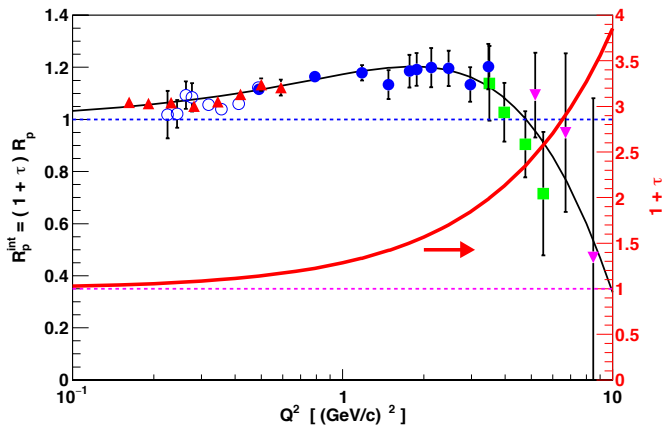}
\caption{Plot of $R_p^{int}$, defined in Eq.(\ref{eq30}), versus $Q^2$.}
\label{fig:2}
\end{figure}

The usual proton form factor ratio is defined as follows using the Sachs form factors%
\begin{equation}
R_{p} \equiv \frac{G_{E}^{p}}{G_{M}^{p}/\mu _{p}}  \label{eq29}
\end{equation}%
and is shown in Fig.~\ref{fig:1};
see \citep{Crawford:2010gv} for
references to the data and to the so-called GKex vector meson based model \citep{Lomon:2006sf} shown as a solid line in the figure. The ratio based on the {\it intrinsic} form factors is then immediately given by%
\begin{equation}
R_{p}^{int}=(1+\tau ) R_{p}  \label{eq30}
\end{equation}%
and is shown in Fig.~\ref{fig:2}.
Note that this has the boost factor squared (going as $1+\tau$ and shown in the
right panel as a red line) rather than just linearly as in the individual form
factors. Clearly this introduces large modifications at high momentum transfers.
Indeed, the {\it intrinsic} results are relatively flat as functions of $Q^2$
and differ from unity by less than roughly 20\%. This is consistent with the
physically reasonable expectation that the proton's intrinsic charge and
magnetization spatial distributions are similar. Finally, we note that the $Q^2$
dependence arising largely from $(1+\tau)$ in our model is the basis for the
observed discrepancy~\citep{Bonner:2017} between the elastic form factors
measured via the Rosenbluth technique and recoil polarization method and widely
understood to be due to two photon contributions to the radiative corrections to
the elastic electron-proton cross section~\citep{Henderson:2016dea}. 

Sachs form factors modified by factors of $(1+\tau)^{\pm n/2}$ have been proposed 
long ago~\citep{Yennie:1957aa} and more recently~\citep{Friar:1997js} but were
never adopted.  However, the particular modification arising from our model
is unique in explaining the measured $Q^2$ dependence of $R_p$ as predominantly a relativistic 
boost effect, as illustrated in Fig.~\ref{fig:2}.

With the Sachs elastic form factors corrected for recoil effects, we are now
ready to correct the RMS charge and magnetic radii using the same procedure; we
call the corrected radii $r_{E,M}^{nr}$ . The usual definition is obtained by
expanding the $j_{0}$ spherical Bessel function for low momentum
transfer to obtain%
\begin{equation}
r_{E}^{scatt}\equiv \sqrt{\frac{3}{2m_{N}^{2}}\left\vert \left( \frac{d}{d\tau
}G_{E}\right) _{\tau \rightarrow 0}\right\vert };  \label{eq31}
\end{equation}%
this is simply a re-writing of Eq.~(\ref{eqadd1}). 
We then proceed to correct for recoil effects using our HO model. For this one
must include the $\sqrt{1+\tau }$ factor in Eq.~(\ref{eq27}),
obtaining at small momentum transfer
\begin{eqnarray}
G_{E}^{st} &=&\left( 1+\frac{1}{2}\tau +\cdots \right) \nonumber\\
 &&\quad\cdot \left( 1-\frac{2}{3}\tau m_{N}^{2}\left[ r_{E}^{scatt}\right] ^{2}+\cdots \right)\\&=&
1-\left[ \frac{2}{3}m_{N}^{2}\left[ r_{E}^{scatt}\right] ^{2}-\frac{1}{2} \right] \tau +\cdots   \\
&\equiv&1-\frac{2}{3}\tau m_{N}^{2}\left[ r_{E}^{nr}\right]^{2}+\cdots \label{eq34}
\end{eqnarray}
and leading to the relationship
\begin{equation}
r_{E}^{nr}=\sqrt{\left[ r_{E}^{scatt}\right] ^{2}-\Delta_N },  \label{eq35}
\end{equation}%
where one has%
\begin{eqnarray}
\Delta _{p} &\equiv &\frac{3}{4m_{p}^{2}}=0.0332\mathrm{\ fm}^{2}
\label{eq35a} \\
\Delta _{n} &\equiv &\frac{3}{4m_{n}^{2}}=0.0331\mathrm{\ fm}^{2}
\label{eq35b}
\end{eqnarray}%
for protons and neutrons, respectively. The same arguments for the
magnetic form factor where the required boost factor is now
$1/\sqrt{1+\tau}$ (see Eq.~(\ref{eq28})) leads to the expression%
\begin{equation}
r_{M}^{nr}=\sqrt{\left[ r_{M}^{rel}\right] ^{2}+\Delta_N }.  \label{eq36}
\end{equation}%
We note that the same result for the proton charge form factor was
obtained in \citep{Giannini:2013bra}, arguing from a very different point of
view: see also~\cite{Licht:1970de}, on which that work is based.
\citep{Robson:2013nwa} also derives the result within a factor of two. 

It is to be noted that the Darwin-Foldy correction to the proton charge radius
discussed in~\citep{Friar:1997js,Jentschura:2010ty} and in the appendix
to~\citep{Krauth:2015nja} has a similar form to $\Delta_N$ but has a very 
different physical origin and is opposite in sign.

It is important to understand that these effects due to recoil do not go away if
electron scattering data are obtained at ever smaller values of the momentum
transfer. As the above expressions clearly show, the relativistic boost factor
arising from $(1+\tau)^{\pm 1/2}$ deviates from unity at order $Q^2$; however,
that is the order needed to extract the charge or magnetic radii. In other words,
being locked together at the same order when expanding in powers of $Q^2$, the
effects can never be separated, no matter how small the momentum transfer
becomes.

Frequently, it is argued~\citep{Pohl:2013yb} that the effective value of $Q \cdot
r_{p}$ is very small for these atomic systems ($\sim 10^{-5}$) and that the
coordinate space and momentum space values converge. However, the above argument
on the ``locking'' of the boost factor with the radius shows that the smallness
of this product is not sufficient to make relativistic and non-relativistic radii
effectively the same. It is not the scale of momentum transfer that is critical
(as long as it is small enough to allow only terms of quadratic order to be
considered), but the fact that a scattering process is analyzed in momentum space
and measurements of an atomic system are studied in coordinate space. 

Specifically, using the Bernauer value for the proton rms charge 
radius~\cite{Bernauer:2010wm},
and the PDG values~\citep{Agashe:2014kda} for the proton magnetic, neutron 
charge, and neutron magnetic rms radii, one has the following:
\begin{eqnarray}
r_{E,p}^{nr} &=& \sqrt{(0.879)^{2}-0.0332}\nonumber\\
&=&0.860\pm 0.008~{\rm  fm}\\
r_{M,p}^{nr} &=&\sqrt{(0.777)^{2}+0.0332}\nonumber\\ 
&=& 0.7981\pm 0.013\pm 0.010~{\rm fm}\\ 
r_{M,n}^{nr} &=&\sqrt{(0.862)^{2}+0.0331}\nonumber\\
&=&0.8810^{+0.009}_{-0.008}~{\rm fm}\\
\left[ r_{E,n}^{nr}\right] ^{2} &=&-0.1161-0.0331\nonumber\\
&=&-0.1492\pm 0.0022~{\rm fm}^{2} \ .
\end{eqnarray}
Here the uncertainties are taken from Bernauer~\cite{Bernauer:2010wm} and from the PDG
compilation~\citep{Agashe:2014kda}, respectively. Note that the electric result for the neutron
is traditionally expressed as the square of the radius, which is
negative.

\begin{figure*}[!ht]
\centering
\includegraphics[viewport=0 0 526 354, width=0.8\textwidth, clip]{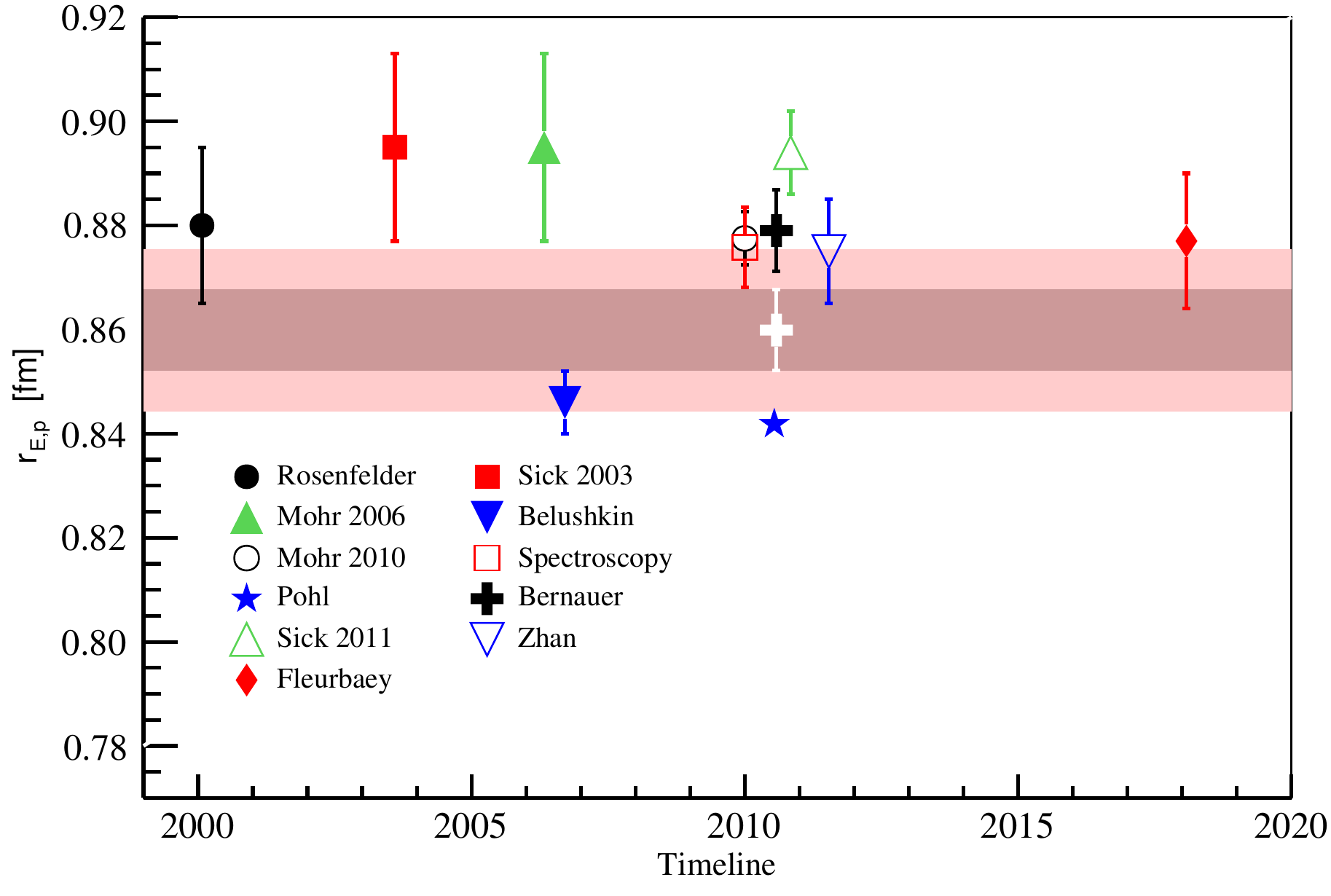}
\caption{Timeline of recent
  determinations~\citep{Rosenfelder:1999cd,Sick:2003gm,Mohr:2008fa,
  Belushkin:2006qa,Mohr:2012tt,Pohl:2013yb,Bernauer:2010wm,
  Sick:2011zz,Zhan:2011ji,Fleurbaey:2018aa} of
  the proton charge radius $r_{E,p}$.  For earlier experimental results see~\cite{Pohl:2010zza}.
  The Bernauer value for the
  proton charge radius, namely $r^{mom}_{E,p} = r^{rel}_{E,p}$, is
  indicated with a black cross lying at 0.879, while its corrected
  value $r^{coord}_{E,p} = r^{nr}_{E,p}$ is indicated by the white
  cross at 0.860. The shaded bands show the 1$\sigma$ and 2$\sigma$
  uncertainties about the latter and may then be compared with the
  Lamb shift value indicated by a star and lying at about 0.84. }
\label{timeline}
\end{figure*}

The proton charge radius discrepancy has arisen from the different
values resulting from a precise determination using the Lamb shift in
muonic hydrogen (0.84087 $\pm$ 0.00026 $\pm$ 0.00029)~\citep{Antognini:1900ns}
disagreeing with the CODATA 2010 value (0.8775 $\pm$
0.0051)~\citep{Mohr:2012tt}, largely determined by the most precise value
resulting from elastic electron proton scattering~\citep{Bernauer:2010wm}.  This
amounts to more than 4\% difference, whereas the stated total uncertainty 
in the Bernauer electron scattering result is quoted as 0.9\%. 
The corrected proton charge radius ($r^{nr}_E$) resulting from the boost 
between Breit and lab. frames
as calculated in our model decreases the electron scattering
determination of $r^{scatt}_{E}$ towards the muonic hydrogen value.  The
resulting discrepancy in the different determinations of $r_{E,p}$ is
halved using the corrected value and now differs from the muonic
Lamb shift value by only about 2\%. Further, the corrected value here
for the proton charge radius is not inconsistent with the value
determined using the hydrogen atom, within experimental uncertainty.
Fig.~\ref{timeline}
shows a selection of determinations of the proton rms charge radius since the year 2000
including all the work cited here.  In particular, it shows the electron scattering determination
from  Bernauer and our corrected value with both 1$\sigma$ and 2$\sigma$ uncertainties.
If one accepts the ideas we have developed in this paper, then one is left with a discrepancy
%to explain and, the remaining discrepancy being commensurate with the
%size 
of order $\alpha$, which could be explained by other model
dependences in the problem (experimental systematic uncertainties,
radiative corrections, {\it etc.})

The $Q^2$ independent recoil correction to the rms charge radius
derived from elastic electron scattering can also be evaluated for the
deuteron, the triton, and the helium isotopes.  For heavier nuclei,
the correction factor above is replaced as follows:
\begin{equation}
\Delta_N = \frac{3}{4m_{N}^{2}}\rightarrow \Delta_N\cdot \left( \frac{m_{N}%
}{m_{A}}\right) ^{2},  \label{eq37}
\end{equation}%
where $m_{A}$ is the mass of that heavier system.
In Table~\ref{tableRp}, 
\begin{table*}
\centering
\begin{tabular}{|r|c|c|c|}
\hline 
Nucleus & $r_{E}$ & $r_{E}$ & $r_{E}$ \\ 
              & e-scattering & e-scattering corrected &   muonic atom \\ 
               & (fm)                 & (fm) & (fm) \\ \hline
$^1$H & 0.879 $\pm$ 0.008~\cite{Bernauer:2010wm} &  0.860 $\pm$ 0.008 & 0.84087 $\pm$ 0.00039 \\
$^2$H & 2.13 $\pm$ 0.01~\cite{Sick:2008aa} &  2.12 $\pm$ 0.01 &
       2.12562 $\pm$ 0.00078 \\
$^3$H  & 1.755 $\pm$ 0.087~\cite{Sick:2008aa} &  1.751 $\pm$ 0.087&  \\
$^3$He & 1.959 $\pm$ 0.034~\cite{Sick:2008aa} & 1.955 $\pm$  0.034 & \\
$^4$He  &  1.680 $\pm$ 0.005~\cite{Sick:2008aa} & 1.678 $\pm$ 0.005 &  \\
\hline
\end{tabular}
\caption{Comparison of charge radii measured by electron scattering
  from various light nuclei and the corrected values following the
  procedure proposed herein.  The muonic determinations for the proton and the deuteron~\cite{Pohl:2016}
  are also given.  We note that high precision values for the helium isotopes 
  will be forthcoming from muonic atom experiments.}
\label{tableRp}
\end{table*}
we summarize the results based on our model.  We note that the
corrected deuteron rms charge radius from electron scattering is in
excellent agreement with the recent high precision value obtained from
measurements of the Lamb shift in muonic deuterium.  We provide
predictions for the rms charge radii of the helium isotopes, which are
being determined to high precision in ongoing experiments that employ
measurement of the Lamb shift in muonic atoms.  New high precision
measurements of the Lamb shift in electronic hydrogen are in progress
and results are expected soon.

We have also considered how the conclusions here can be validated by
experiment.  We point out that the correction we derive cannot be
separated by going to lower $Q^2$ in electron scattering experiments
nor by comparison of elastic electron and muon scattering on the proton.
However, based on the estimates made here, a precision comparison of
electron scattering from the proton with electron scattering from the
deuteron at low $Q^2$ should deviate at the level of about 1.5\%, since the
corrections we derive differ at this level.  For highest precision,
such measurements should be carried out using the same apparatus and
systematics must be minimized.  We note that internal radiative
corrections, which arise mainly from the incident and scattered
electrons, should be quite similar for proton and deuteron targets.

In summary, we argue that corrections between the Breit and lab.
frames are important in interpreting the form factors of the nucleon
as determined in relativistic electron scattering.  We have
constructed a model to estimate these corrections.  In this 
model, the observed significant decrease of the ratio of the proton
elastic form factors as a function of $Q^2$ is understood as a
predominantly relativistic effect.  Furthermore, in this model,
the proton charge radius as determined in electron scattering has a
correction that reduces its value towards that resulting from the
precision determination using the Lamb shift in muonic hydrogen. The
model provides predictions for the rms charge radii of the deuteron, triton,
and helium isotopes.

The work here underlines the importance of having
a model of the nucleon that allows boosting between different reference frames.
For example, in models where relativistic quarks are
confined via a ``bag" one must confront the problem of how to boost
the latter in going between the frames that inevitably enter in
electron scattering. One possible, but different, related case that
could be studied to test the ideas is that of relativistic (covariant,
boostable) modeling of the deuteron~\citep {VanOrden:2016aa}.  There, 
one could directly compute the elastic form factor, while in parallel
also computing the ground-state charge distribution and
then Fourier transforming it to momentum space.  Upon comparing the two
results it is likely that differences will be found that relate to the
boost issues raised in the present study.

%\begin{acknowledgement}
We thank Jan Bernauer for valuable discussion and Randolf Pohl and Don Robson 
for helpful communications. The authors' research is supported by the Office of
Nuclear Physics of the U.S. Department of Energy under grant Contract Numbers
DE-SC0011090 and DE-FG02-94ER40818.
%\end{acknowledgement}

\bibliographystyle{mybibstyle} 
\bibliography{FormFactor-epj-rev}

\end{document}